\documentclass[letterpaper, 10 pt, conference]{ieeeconf}  % Comment this line out if you need a4paper

\usepackage{graphicx}
\IEEEoverridecommandlockouts 
\usepackage{amsmath,amssymb}
\usepackage{mathrsfs}
\usepackage{amsfonts}
\usepackage{tikz}
\usetikzlibrary{automata,positioning}
\usepackage{color}
\usepackage{algorithm,algorithmicx,algpseudocode}
\usepackage{graphicx}
\let\labelindent\relax
\usepackage{enumitem}

\newcommand\mynodedistance{1.5cm}

\tikzset{every state/.style={minimum size=0.5pt}}
\tikzset{every edge/.append style={font=\small}}

\newcommand{\blue}[1]{{\color{blue}  #1}}

\newcommand{\TA}{\mathcal{A}}
\newcommand{\tree}{\mathcal{E}}
\newcommand{\LTS}{\mathcal{T}}
\algnewcommand\Or{\textbf{or}}
\algnewcommand\algorithmicforeach{\textbf{for each}}
\algdef{S}[FOR]{ForEach}[1]{\algorithmicforeach\ #1\ \algorithmicdo}
\usepackage{stmaryrd}

\newcommand{\RUNS}[1]{\ensuremath{\llbracket#1\rrbracket}}

\newtheorem{remark}{\bf Remark}[section]
\newtheorem{proposition}{\bf Proposition}[section]
\newtheorem{assumption}{\bf Assumption}[section]
\newtheorem{definition}{\bf Definition}[section]
\newtheorem{example}{\bf Example}[section]

\newtheorem{problem}{\bf Problem}[section]

\begin{document}
\title{\LARGE \bf
Control Synthesis for Parametric Timed Automata under Unavoidability Specifications}

\author{Ebru Aydin Gol% <-this % stops a space
\thanks{This work has received funding from the European Union's Horizon 2020 research and innovation programme under the Marie Sk\l{}odowska-Curie grant agreement No 798482.}
\thanks{Ebru Aydin Gol is with the Department of Computer Engineering,
      Middle East Technical University, Ankara/TURKEY
        {\tt\small \{ebrugol\}}{\tt\small @metu.edu.tr}}%
        }

\maketitle              % typeset the header of the contribution
\begin{abstract}
Timed automata (TA) is used for modeling systems with timing aspects.  
A TA extends a finite automaton with a set of real valued variables called clocks, that measure the time and constraints over the clocks guard the transitions. A parametric TA (PTA) is a TA extension that allows parameters in clock constraints. In this paper, we focus on synthesis of a control strategy and parameter valuation for a PTA such that each run of the resulting TA reaches a target location within the given amount of time while avoiding unsafe locations. 
We propose an algorithm based on depth first analysis combined with an iterative feasibility check. 
The algorithm iteratively constructs a symbolic representation of the possible solutions, and employs a feasibility check to terminate the exploration along infeasible directions. Once the construction is completed, a mixed integer linear program is solved for each candidate strategy to generate a parameter valuation and a control strategy pair.  We present a robotic planning example to motivate the problem and to illustrate the results. 
\end{abstract}

\section{Introduction}\label{intro}

Timed automata (TA)~\cite{alur1994theory} is used for modeling systems with timing aspects.  A TA extends a finite automaton with a set of real valued variables called clocks that measure the time. The clocks enrich the semantics, and the constraints over the clocks restrict the behavior of the automaton.  The examples of real-time systems modeled as timed automata includes rail-road crossing systems~\cite{Wang:2004}, scheduling problems~\cite{david2009model}, and pace-makers~\cite{pacemakers:2015,10.1007/s10009-013-0289-7}.

The correctness of a TA against high level specifications such as safety, reachability and unavoidability can be verified via model checking algorithms that are implemented in off-the-shelve tools such as UPPAAL~\cite{UPPAAL4:0} and HyTech~\cite{hytech}. A reachability specification requires existence of an execution that reach a target set, whereas, an unavoidability (inevitability) specification requires each execution to reach a target set. 
Using a model checker to verify such a property requires a complete TA model, and designing it for a complex system (or problem) is a very challenging task. Parametric timed automata (PTA) simplifies the design problem by allowing the use of parameters in place of the numeric constants. 
Then, the model generation is completed via parameter synthesis: find a parameter valuation such that the resulting model satisfies the specification~\cite{6895298}. However, parameter synthesis problems are, in general, undecidable~\cite{Etienne2019_decidable}.

The control of timed automata problem deals with the synthesis of a controller that monitors and affects the behavior of the timed automata such that the resulting controlled system satisfies the specification. In literature~\cite{ASARIN1998447,10.1007/978-3-540-45069-6_18,10.1007/11539452_9,uppaal-tiga}, the timed automaton is assumed to have controllable and uncontrollable inputs (transitions), and a control strategy that restricts the controllable transitions by both assigning input symbols and delay values is synthesized. In the pioneering work~\cite{ASARIN1998447}, the authors restricted the transitions of a timed automaton by solving a turn-based timed game such that the resulting automaton satisfies a safety property (avoids ``bad" states). An on-the-fly algorithm for safety and reachability specifications is developed in~\cite{10.1007/11539452_9,uppaal-tiga} to generate a feedback controller that assigns a control input or a delay value to partial runs. In~\cite{10.1007/978-3-540-45069-6_18}, a controller in the form of a timed transition system is synthesized for partially observable timed automata. %, i.e., locations, controllable actions, a subset of the clocks and a subset of the uncontrollable actions are visible to the controller.
A template-based controller synthesis method for safety specifications is studied in~\cite{templateControl2012}. 

In this paper, we study the problem of synthesizing a control strategy and a parameter valuation pair for a PTA such that the resulting TA satisfies an unavoidability specification. In particular, we require each run to reach a set of target locations ($L_T$) within a given amount of time ($D$) while avoiding unsafe locations ($L_A$). 
We consider control strategies that map a TA path (sequence of locations and transitions) to an input and a delay value pair. %As a result, the strategy can be represented as a TA. 
It is important to note that the controlled TA can be non-deterministic. Thus, it is necessary to ensure that each possible run satisfies the constrained unavoidability specification. 
To solve this problem, we represent candidate strategies symbolically as a tree with respect to the specification.   Then, we employ a Mixed Integer Linear Programming (MILP) to generate a control strategy and a parameter valuation pair from a symbolic tree. Furthermore, we present an efficient algorithm to construct the candidate solutions (trees). The algorithm constructs the candidate trees in a depth first manner and employs an MILP based feasibility check to terminate the exploration along the infeasible directions. Finally, we show that the algorithm is complete under a mild non-zeno assumption~\cite{ASARIN1998447}.

 As summarized, in general, the parameter and controller synthesis problems are studied separately. Here, we tune parameters and restrict transitions via controller synthesis such that the resulting automaton satisfies a specification, thus we combine both problems for constrained unavoidability specifications. Parameter and controller synthesis is previously studied under safety~\cite{7842500}  and reachability~\cite{GOLReach}. In~\cite{7842500}, a symbolic parameter synthesis method is extended to incorporate symbolic constraints over the TA inputs, whereas in~\cite{GOLReach}, a path is searched in a depth first manner with an MILP  encoding to find parameters.
 % In~\cite{templateControl2012}, the template definition reduces the controller synthesis problem to boolean parameter synthesis problem. 

\section{Preliminaries}\label{prelim}

% \textbf{Notation:} $\mathbb{N}$ is the set of natural numbers, $\mathbb{R}_{\geq 0}$ is the set of non-negative real numbers. 

\subsection{Timed Automata}

A timed automaton (TA)~\cite{alur1994theory} is a finite-state machine extended with a finite set of real-valued clocks denoted by $C$. A clock $x\in C$ measures the time spent after its last reset. Clock constraints define timed conditions for transitions (guard). A clock constraint is defined with the following grammar $\phi := x \sim c \mid \phi \wedge \phi$ where $x \in C$ is a clock, $c \in \mathbb{N}$ is a constant and $\sim \in \{<,\leq, >, \geq\}$ ($\mathbb{N}$ is the set of natural numbers). A constraint is called parametric if some of the numeric constants are represented with parameters. The set of clock constraints over $C$ is defined as $\Phi(C)$. For a parametric clock constraint $\phi$ with $P$ as its set of parameters and a parameter valuation $\gamma: P \to \mathbb{N}$, $\phi(\gamma)$ is the constraint obtained by replacing parameters in $\phi$  with the corresponding constants from $\gamma$, e.g, for $\phi = x > p_1 \wedge y \leq p_2$, and valuation $\gamma(p_1) = 3, \gamma(p_2) = 4$, $\phi(\gamma) = x > 3 \wedge y \leq 4$.

A clock valuation $v: C \to \mathbb{R}_{\geq 0}$ assigns non-negative real values to each clock. The notation $v \models \phi$ denotes that the clock constraint $\phi$ evaluates to true when each clock $x$ is replaced with the corresponding valuation $v(x)$.  Two operations are defined for clock valuations: delay and reset. For a clock valuation $v$ and $ d \in \mathbb{R}_{\geq 0}$, $v+d$ is the clock valuation obtained by adding $d$ to each clock, i.e., $(v+d)(x) = v(x) +d $ for each $x\in C$. For $\lambda \subseteq C$, $v[\lambda]$ is the clock valuation obtained after resetting each clock from $\lambda$, i.e.,   $v[\lambda](x) = 0$ for each $x\in \lambda$ and $v[\lambda](x) = v(x)$ for each $x\in C\setminus \lambda$.

% TODO SHORTEN NO BULLET 
\begin{definition}[Timed Automata]
A \textit{timed automaton}  $\TA = (L, l_0, \Sigma, C, \Delta )$ is a tuple, where $L$ is a finite set of locations, $l_{0} \in L$ is the initial location, $\Sigma$ is a finite input alphabet, $C$ is a finite set of clocks and $\Delta \subseteq L \times \Sigma \times 2^C \times  \Phi(C) \times L$ is a finite transition relation. 
% \begin{itemize}
% \item $L$ is a finite set of locations, 
% \item $l_{0} \in L$ is the initial location, 
% $\Sigma$ is a finite input alphabet, 
% \item $C$ is a finite set of clocks, 
% \item $\Delta \subseteq L \times \Sigma \times 2^C \times  \Phi(C) \times L$ is a finite transition relation, and 
% %\item $Inv: L \to \Phi(C)$ is an invariant function.
% \end{itemize}
\end{definition}
\noindent
For a  transition $e = (l_s, a, \lambda, \phi, l_t) \in \Delta$, $l_s$ is the source location, $l_t$ is the target location, $a\in\Sigma$ is the input symbol, $\lambda$ is the set of clocks reset on $e$ and $\phi$ is the guard tested for enabling $e$. The set of enabled input symbols in a location $l$ is denoted by $\Sigma(l) = \{a \mid (l, a, \lambda, \phi, l') \in \Delta \}$. % \footnote{The classical definition includes an invariant function for locations. It is known that a TA can be encoded as an equivalent one without invariants.}  
The set of locations that can be reached from $l$ under input $a$ is defined as $Post(l, a) = \{ l' \mid  (l, a, \lambda, \phi, l') \in \Delta\}$.
A clock $x_0 \in C$ is used to measure the time passed since the start of the execution, thus it is not reset on any transition of $\TA$.
%  \text{ for some } \lambda, \phi, l'

A TA is called parametric (PTA) if it contains a parametric  clock constraint. Given a PTA $\TA$ with a set of parameters $P$ and a valuation $\gamma: P \to \mathbb{N}$ for its parameters, $\TA(\nu)$ is the TA obtained by replacing each parameter with the corresponding constant from the valuation $\gamma$. 

The semantics of a TA is given by a timed transition system (TTS). An TTS is a tuple $\LTS = (S, s_0, \Gamma, \to)$, where $S$ is a set of states, $s_0 \in S$ is an initial state, $\Gamma$ is a set of symbols, and $\to \subseteq S \times \Gamma \times S$ is a transition relation. A transition $(s,a,s') \in \to$ is also shown as $s \stackrel{a}{\to} s'$.

\begin{definition}[TTS semantics for TA]\label{defn:lts} Given a timed automaton $\TA = (L, l_0, \Sigma, C, \Delta )$, the timed transition system $\LTS(\TA)= (S, s_0, \Gamma, \to)$ is defined as follows:
\begin{itemize}
\item $S = \{(l, v) \mid l \in L, v\in \mathbb{R}_{\geq 0}^{|C|}\}$,
\item $s_0 = (l_0, \textbf{0})$, where $ \textbf{0}(x) = 0$ for each $x\in C$, 
\item $\Gamma = \Sigma \cup  \mathbb{R}_{\geq 0}$, and the transition relation defined by the following rules:
\begin{itemize}
\item delay transition: $(l, v) \stackrel{d}{\to} (l, v+d)$ if $v+d \models Inv(l)$
\item discrete transition: $(l, v) \stackrel{a}{\to} (l', v')$ if there exists $(l,a, \lambda, \phi, l') \in \Delta$ such that $v \models \phi$, and $v' = v[\lambda]$. % , and $v' \models Inv(l')$.
\end{itemize}
\end{itemize}
\end{definition}
% The notation $s  \stackrel{a}{\to}_d s'$ is used to denote a delay transition of duration $d$ followed by a discrete transition over input symbol $a$ from $s$ to $s'$, i.e., $s  \stackrel{d}{\to} s  \stackrel{a}{\to} s'$.  
A run $\rho$ of $\TA$ is an alternating sequence of delay and discrete transitions:
%\begin{equation}\label{eq:run}
%\rho: s_0 \stackrel{a_0}{\to}_{d_0} s_1 \stackrel{a_1}{\to}_{d_1} s_2 \stackrel{a_2}{\to}_{d_2} \ldots,
%\end{equation}
\begin{equation}\label{eq:run}
\rho: (l_0, v_0) \stackrel{d_0}{\to}  (l_0,v_0 + d_0) \stackrel{a_0}{\to} (l_1, v_1) \stackrel{d_1}{\to}  \ldots, % (l_1, v_1+d_1)  \stackrel{a_1}{\to},
\end{equation}
where $v_0$ is $\textbf{0}$, $a_i \in \Sigma$ and $d_i \in \mathbb{R}_{\geq 0}$ for each $i \geq 0$.
A run is called \textit{maximal} if it is either infinite or can not be extended by a discrete transition. 
The set of all runs of $\TA$ is denoted by $\RUNS{\TA}$. A path $\pi$ of $\TA$ is an interleaving sequence of locations and transitions, $\pi: l_0, e_1, l_1, e_2, \ldots$.  A path $\pi$ is \textit{realizable} if  there exists a delay sequence $d_0, d_1, \ldots$ such that $(l_0, v_0) \stackrel{d_0}{\to}  (l_0,v_0 + d_0) \stackrel{a_0}{\to} (l_1, v_1) \stackrel{d_1}{\to}  \ldots$ is a run of $\TA$, and for every $i \geq 1$, the $i$th discrete transition is taken according to $e_i$, i.e., $e_i = (l_{i-1}, a_{i-1}, \lambda_{i-1}, \phi_{i-1}, l_i)$, $v_{i-1} + d_{i-1} \models \phi_{i-1}$,  and $v_i = (v_{i-1} + d_{i-1})[\lambda_{i-1}]$.

In this work, we study control strategies that assign a delay value and an input symbol to a finite path:
\begin{definition}[Control Strategy]\label{def:strategy}
A control strategy $\mathcal{C}: (L \times \Delta)^n \times L \to  \mathbb{R}_{\geq 0} \times \Sigma$, $n \geq 0$, for a TTS $\LTS(\mathcal{A})= (S, s_0, \Gamma, \to)$ of a TA  $\TA$ (Defn.~\ref{defn:lts}) maps a path $\pi$ of $\TA$ to a delay and input symbol pair.
 A run $\rho$ as in~\eqref{eq:run} is generated in closed loop with a strategy $\mathcal{C}$ if for each $n \geq 0$: 
 \begin{enumerate}[label=(\alph*)]
 \item $ \mathcal{C}(l_0, e_1, \ldots, l_n) =  (d_n, a_n)$, 
 \item there exists $ e_{n+1} = (l_n, a_n, \lambda_n, \phi_n, l_{n+1}) \in \Delta$ such that $v_{n} + d_n \models \phi_n$, and $v_{n+1} = (v_{n} + d_n)[\lambda_n]$.
 \end{enumerate}
 \end{definition}
For a timed automaton $\TA$ and a valid strategy $\mathcal{C}$ for $\TA$, the set of all runs of $\TA$ that is generated in closed loop with $\mathcal{C}$ is denoted by $\RUNS{\TA_{\mathcal{C}}}$. A strategy only limits the transitions of $\TA$, thus $\RUNS{\TA_{\mathcal{C}}} \subseteq \RUNS{ \TA }$. Note that the resulting controlled TA can be non-deterministic since there can be multiple transitions satisfying  condition-(b) from Def.~\ref{def:strategy}

\begin{definition}[$(L_T, L_A, D)$-satisfaction] Let $\TA = (L, l_0, \Sigma, C, \Delta )$ be a timed automaton, $L_T \subset L$ and $L_A \subset L$ be subsets of its locations, and $D \in \mathbb{N}$ be a time bound. 
A run $\rho$ %= (l_0, \mathbf{0})  \stackrel{a_0}{\to}_{d_0} (l_1, v_1) \stackrel{a_1}{\to}_{d_1} (l_2, v_2)  \stackrel{a_0}{\to}_{d_0} \ldots $
as in~\eqref{eq:run}  of $\TA$ satisfies the reach-avoid specification with deadline $(L_T, L_A, D)$ if there exists $i \in \mathbb{N}$ such that $l_i \in L_{T}$, $l_j \not \in L_{A}$ for each $j < i$, and $v_i(x_0) \leq D$.
\end{definition}
\begin{remark}
This specification can be expressed as a temporal logic formula with bounded until operator ($\neg L_A \mathtt{U}_{[0, D]}  L_T$). As we focus on this particular specification, further details on the syntax and semantics of temporal logics are not included. Furthermore, an alternative way to enforce the deadline is to add $x_0 \leq D$ to each transition that ends in a location $ l \in L_{T}$. As our goal is to enforce the overall specification via controller and parameter synthesis, we integrate this to the specification instead of the TA.
\end{remark}

\section{Problem Formulation}\label{problem}

\begin{problem}\label{prob:main} Given a PTA $\TA = (L, l_0, \Sigma, C, \Delta )$ with parameter set $P$, an interval $I_p \subset \mathbb{N}$ for each parameter $p \in P$, and a reach-avoid specification $(L_T, L_A, D)$, find a parameter valuation $\gamma: P \to \Pi_{p \in P} I_p$ and a feedback control strategy $\mathcal{C}
$ as in Defn.~\ref{def:strategy} such that each run $\rho \in \RUNS{\TA_{\mathcal{C}}({\gamma})}$ satisfies $(L_T, L_A, D)$.
\end{problem}

Intuitively, our goal is to find a parameter valuation $\gamma$, and restrict the behaviors of $\TA(\gamma)$ via controller synthesis, such that each remaining run reaches $L_T$ within $D$ time units while avoiding $L_A$.
Our solution for this problem constructs a symbolic exploration tree for the given PTA. Central to the proposed method is the iterative construction of the symbolic model equipped with a MILP based feasibility analysis guided by the specification. This approach avoids computation of symbolic states that can not be part of the solution, i.e., not reachable by a TA $\TA_\mathcal{C}(\gamma)$  solving Prop.~\ref{prob:main}.

The developed method is presented for PTA satisfying the following assumption. The extension of the method to TA violating the assumption is explained in Remark~\ref{remark:samephi}. 

\begin{assumption}\label{assm:samephi} For a TA $\TA = (L, l_0, \Sigma, C, \Delta )$
\[ \text{if } (l_s, a, \lambda, \phi, l_t),  (l_s, a, \lambda', \phi', l_t')  \in \Delta, \text{ then } \phi = \phi'.\]
\end{assumption}

The assumption states that the guards of all transitions that leave the same state ($l_s$) under the same input ($a$) are the same ($\phi = \phi'$). The following example illustrates Prob.~\ref{prob:main} over a time-constrained task planning problem for a robot. 

\begin{example}\label{ex:running}
An example timed automaton is shown in Fig.~\ref{fig:example}. The automaton represents a task planning problem for a robot. The robot has three tasks $a$, $b$ and $c$. It needs to complete either $a$ or $b$ and then $c$. Each task is represented with a location in the timed automata ($l_a$, $l_b$, $l_c$). In addition, it is assumed that the machines (tools/room) that the robot needs for a task can be busy. In this case, the robot waits for at least $p_1$ time units (locations $l'_{a}$, $l'_{b}$, $l'_{c}$). Thus, when the robot decides to perform a task, say $a$, it either (1) reaches location $l'_{a}$, and then it can move to $l_a$, or (2) it reaches $l_a$ without waiting. The other tasks are represented similarly. The task durations have relative constraints. For example, the bound for the duration of task $c$ should be ``more than two times and less than three times" of the bounds defined for the duration of task $a$. These relative constraints are captured with the parametric constraints. The parameter intervals are $I_{p_1} = I_{p_2} = \{2,3,4\}$. Further details are given in Fig.~\ref{fig:example}. The input alphabet of the TA is $\Sigma = \{a,b,c,d\}$. The goal is to generate a strategy $\mathcal{C}$ and a parameter valuation $\gamma$ for $p_1$ and $p_2$ such that each run  $\rho \in \RUNS{\mathcal{A}_{\mathcal{C}}(\gamma)}$ reaches $l_t$ in $15$ time units without visiting $l_{d}$, i.e, the specification is $(L_T,L_A,D)$ with $L_T = \{l_t\}$, $L_A = \{l_d\}$, and $D = 15$.
\end{example}

\begin{figure}[h]
\vspace{-0.4cm}
\begin{center}
\begin{tikzpicture}[shorten >=1 pt, node distance = \mynodedistance, on
grid, auto]
\tikzstyle{every node}=[font=\scriptsize]
%\node [state, initial,initial where=above, label = {below: $test$}] (l_0) {$l_0$};
\node [state] (l_0) {$l_0$};
\node [state] (l_wa) [above right = of l_0] {$l'_{a}$};
\node [state] (l_a) [above right = of l_wa, yshift=-0.5cm, xshift=0.5cm] {$l_a$};
\node [state] (l_wc) [right= of l_0, xshift=1.2cm] {$l'_c$};
\node [state] (l_c) [right= of l_wc, xshift=1cm] {$l_c$};
\node [state] (l_t) [right= of l_c, xshift=1cm] {$l_t$};
\node [state] (l_wb) [below right = of l_0] {$l'_{b}$};
\node [state] (l_b) [below right = of l_wb, yshift=0.5cm, xshift=0.5cm] {$l_b$};
\node [state] (l_cc) [right= of l_a, xshift=1cm] {$l_{d}$};
\path [->]
(l_0) edge [bend left ] node [right, pos=0.25]{$\blue{a}, \{x\}$}  (l_wa) 
% (l_0) edge [bend left ] node [right, pos=0.1] {$a$}  (l_wa) 

(l_0) edge [bend left=60 ] node {$\blue{a}, \{x,y\}$}  (l_a) 
% (l_0) edge [bend left=60 ] node [left, pos=0.25] {$a$}  (l_a) 

(l_wa) edge [bend left] node [right, pos=-0.05]{$\blue{a},  x \geq p_1$}  (l_a) 
(l_wa) edge [bend left] node [right, pos=0.25]{$ \{x,y\}$}  (l_a) 

(l_a) edge node [right, pos=0.2]{$\blue{c}, \phi_a$}  (l_wc) 
(l_a) edge node [right, pos=0.6]{$ \{x\} $}  (l_wc)

(l_a) edge [bend left=20] node [right, pos=0.4]{$\blue{c},  \phi_a $}  (l_c) 
(l_a) edge [bend left=20]  node [right, pos=0.7]{$\{x\}$}  (l_c) 

(l_wc) edge node [above, pos=0.4]{ $\blue{c}, x \geq  p_1, \{x\} $}  (l_c) 

(l_a) edge node [above, pos=0.4]{$\{x\}, \blue{d}, \phi_d$}  (l_cc) 
(l_a) edge [bend left=60] node [above, pos=0.4]{$\blue{d}, \phi_d $}  (l_t) 

(l_cc) edge node [right, pos=0.2]{$\blue{d},  \phi_c  $}  (l_t) 

(l_c) edge node [above, pos=0.4]{$\blue{c}, \phi_c$}  (l_t)

(l_0) edge [bend right=60 ] node [left, pos=0.5] {$\blue{b}, \{x,y\}$}  (l_b) 
(l_0) edge [bend right ] node [right, pos=0.25]{$\blue{b}, \{x\}$}  (l_wb) 

(l_wb) edge [bend right] node [right, pos=-0.05]{$\blue{b},  x \geq p_1$}  (l_b) 
(l_wb) edge [bend right] node [right, pos=0.25]{$ \{x,y\}$}  (l_b)

(l_b) edge node [right, pos=0.2]{$\blue{c}, \phi_b$}  (l_wc) 
(l_b) edge node [right, pos=0.6]{$ \{x\} $}  (l_wc) 

(l_b) edge [bend right=20] node [right, pos=0.4]{$\blue{c},  \phi_b $}  (l_c) 
(l_b) edge [bend right=20]  node [right, pos=0.7]{$\{x\}$}  (l_c) 
;
\end{tikzpicture}
\end{center}
\vspace{-0.9cm}
\caption{The timed automaton from Ex.~\ref{ex:running}. $l_0$ is the initial location. The control inputs, reset sets and the constraints are shown next to the transitions. For example, the transition from $l'_c$ to $l_c$ is $ (l'_c, \blue{c}, \{x\}, x\geq 4, l_c)$.  The parametric constraints are $\phi_a := p_2 \leq x \wedge x \leq p_1$, $\phi_b := x \geq 5 p_1$, $\phi_c := 2p_1 \leq x \wedge x \leq 3 p_2 \wedge y \geq 12 $, and $\phi_d := x \geq  p_2 \wedge y \geq 12 $. }\label{fig:example}
\vspace{-0.6cm}
\end{figure}

\section{Control and Parameter Synthesis}\label{sec:main}
In this section, we present the proposed method to solve Prob.~\ref{prob:main}, and prove the correctness of the result. The method first constructs an exploration tree that symbolically represents the TA runs, and then solves an optimization problem for each candidate solution (a sub-tree) represented in the tree. We first formally define the exploration tree, and the associated candidate solutions with respect to the specification $(L_T, L_A, D)$. Then, we present an algorithm to synthesize a control strategy-parameter valuation pair without constructing the whole tree, which can be infinite.

\begin{definition}[Exploration Tree]\label{defn:expTree} The exploration tree of a $PTA$ $\TA=(L, l_0, \Sigma, C, \Delta )$ is denoted by $\tree(\TA)$ and it is a rooted tree defined in the following way:
\begin{itemize}[noitemsep,topsep=0pt]
\item The root $r$ is labelled by the initial location $l_0$.
\item If $m \in \tree(\TA)$ is a tree node labelled by $l \in L$, then for each $a \in \Sigma(l)$, and for each $(l, a, \lambda, \phi, l') \in \Delta$ there exists a node $m' \in \tree(\TA)$ that is labelled by $l'$ and is an $a$-successor of $m$.
\end{itemize}
The label and the set of $a$-successors of a node $m$ are denoted by $m(l)$ and $\tree(\TA, m,a)$, respectively.
\end{definition}

An exploration tree characterizes all possible paths of $\TA$. If $\TA$ includes a cycle, i.e., if it has a path $\pi = l_0, e_1, l_1, e_2, \ldots, $ with $l_i = l_j$ for some $i \neq j$, then $\tree(\TA)$ is infinite.  
By the tree definition, there is a one-to-one mapping between a tree path from root to a node and an automaton path. 
Given a node $m \in \tree(\TA)$, the path from root $r$ to $m$ is uniquely defined as $\pi_{r \to m} = m_0,\ldots, m_n$ where $m_0$ is $r$, $m_n$ is $m$, and for each $i=0,\ldots,n-1$ there exists $a_i \in \Sigma$ such that $m_{i+1} \in \tree(\TA, m_{i},a_{i})$.  The corresponding automaton path is  $\pi^\TA_{r \to m} = l_0, e_1, l_1, e_2, \ldots, l_n$ where for each $i = 0,\ldots,n$, $l_i = m_i(l)$, and for each $i=1,\ldots,n$, $e_i =  (l_{i-1}, a_i, \lambda_i, \phi_i, l_i) \in \Delta$ for some $\lambda_i$ and $\phi_i$ ($a_i$ is as in $\pi_{r \to m}$). Before introducing sub-trees characterizing control strategies, we present an MILP based method to decide whether a path is realizable within the given time limit $D$. This method is extended to sub-trees for controller synthesis.

\begin{proposition}\label{prop:milp} Let $\TA=(L, l_0, \Sigma, C, \Delta)$ be a parametric timed automaton with parameter set $P$,  $\{I_p\}_{p \in P}$ be the set of parameter ranges, and $\pi = l_0, e_1, l_1, e_2, \ldots, l_n $ be a path of $\TA$. There exists a parameter valuation $\gamma$  such that $\pi$ is realizable on $\TA(\gamma)$ within $D$ time units if and only if MILP~\eqref{eq:LP}  with the decision variables $\gamma_p$, $p \in P$ and $d_0, \dots, d_{n-1}$ is feasible.
\begin{subequations}\label{eq:LP}
\begin{align} 
&  \gamma_p \in I_p & \text{ for each } p \in P \text{ and }   \\
&  d_i \in \mathbb{R}_{\geq 0}  &  \text{ for each } i=0,\ldots,n-1 \label{eq:LP2} \\
& M(x, \pi, i) \sim g(c)  &   \text{ for each } i = 1, \ldots, n, \nonumber \\
& &  \text{ and for each } x \sim c \text{  from }  \phi_i \label{lp:guard} \\
&  \sum_{i=0}^{n-1} d_i \leq D,& \label{eq:deadline1}
\end{align}
\end{subequations}
where $g(c)$ is $\gamma_p$ if $c$ is parameter $p$, otherwise, i.e., if $c \in \mathbb{N}$, $g(c) = c$, and\begin{align}\label{eq:transform}
M(x, \pi, i) &= d_k + d_{k+1} + \ldots  + d_{i-1}  \text{ and }  \\
& k = \max(\{ m  \mid x \in \lambda_m, m < i\} \cup \{ 0 \}). \nonumber
\end{align}
\end{proposition}
The value of a clock $x$ on a particular transition of $\pi$ is represented as the sum of the delay variables since the last reset of $x$ via $M(\cdot)$~\eqref{eq:transform}. In particular, clock $x$ equals to $M(x, \pi, i)$ on the i-th transition $e_i$ along $\pi$.  

\begin{example}~\label{ex:pathMILP} Consider  the TA introduced in Ex.~\ref{ex:running} and its path $\pi_1 = l_0, l_b', l_b, l_c', l_c$ (edges are omitted for brevity).  Delay values $d_0$, $d_1$, $d_2$, $d_3$, are the positive real valued variables and parameters $\gamma_{p_1}$ and $\gamma_{p_2}$ are the integer valued variables (with domain $\{2,3,4\}$) of the corresponding MILP~\eqref{eq:LP}. The MILP constraints are $C_1 : d_1 - \gamma_{p_1} \geq 0$, $C_2: d_2 - 5\gamma_{p_1} \geq 0$, $C_3: d_3 - \gamma_{p_1} \geq 0$ (from~\eqref{lp:guard}),  and $C_4: d_0 + d_1 + d_2 + d_3  \leq 15$~\eqref{eq:deadline1}. This MILP is feasible. Now, consider the extended path $\pi_2 = l_0, l_b', l_b, l_c', l_c, l_t$. It has an additional delay variable $d_4$. Its constraints are $C_1, C_2, C_3$ as in $\pi_1$ and  $C_5: d_4 - 2\gamma_{p_1} \geq 0$, $C_6: -d_4 + 3\gamma_{p_2} \geq 0$, $C_7: d_2 + d_3 + d_4 \geq 0$ and $C_8: d_0 + d_1 + d_2 + d_3 + d_4 \leq 15$. In this case, the MILP is infeasible. 
\end{example}

\begin{definition}[Proper Sub-tree]\label{defn:expTreeSoln}
A proper sub-tree $\bar \tree(\TA)$ of an exploration tree $\tree(\TA)$ with respect to $(L_T, L_A, D)$ has the following properties
\begin{enumerate}[leftmargin=*,noitemsep,topsep=0pt]
\item The root $r$ of $\bar \tree(\TA)$ is labelled by the initial location $l_0$.
\item For each node $m \in \bar \tree(\TA)$,  $m$ is also node of $\tree(\TA)$, and
\begin{enumerate}[leftmargin=*]
\item $m(l) \in L \setminus L_A$,
\item if $m(l) \in L_T$, then $m$ does not have a successor,
\item if $m(l) \not \in L_T$, then there exists $a^m \in \Sigma(m(l))$, such that for each $(l,a^m,\lambda, \phi,l') \in \Delta$ there exists $m' \in \bar \tree(\TA, m, a^m)$ with $m'(l) = l'$, and for each 
$b \neq a^m$, $\bar \tree(\TA, m,b) = \emptyset$.
\item if $m$ is not root, there is $m' \in \bar \tree(\TA)$ such that $m \in \tree(\TA, m',a)$.
\end{enumerate}
\end{enumerate}
\end{definition}
The proper sub-tree definition ensures that locations from the avoid set $L_A$ are not included in the tree (a), the leaf nodes are labelled by the target locations ($L_T$) (b), a unique input $a \in \Sigma$ is assigned to each internal node (non-leaf) and each location that is reachable under the assigned input is represented by the corresponding nodes (c), and the tree is connected (d).  A proper sub-tree symbolically characterizes a candidate solution in terms of an input assignment, and integrates specifications $L_T$ and $L_A$. 

Next, we define a control strategy $\mathcal{C}$ and a parameter valuation $\gamma$ from a proper sub-tree $\bar \tree(\TA)$ by solving a MILP over $\{\gamma_p \mid p \in  P \}$ and $\{d_m \mid  m \in Int(\bar \tree(\TA) )\}$, where $Int(\bar \tree(\TA) )$ is the set of internal (non-leaf) nodes of $\bar \tree(\TA)$.
\begin{subequations}\label{eq:LPTree}
\begin{align} 
& \gamma_p \in I_p   & \text{ for each } p \in P \text{ and }  \\
&  d_{m} \in \mathbb{R}_{\geq 0}   & \text{ for each } m  \in Int(\bar \tree(\TA)) \\
&  M^\tree(x, m') \sim g(c)  &  \text{ for each } m  \in Int(\bar \tree(\TA))  \nonumber \\
&  &  \text{ and for each } x \sim c \text{  from }  m^\phi \label{eq:guardtree} \\
&   \sum_{m \in \pi_{r \to m_t}} d_m \leq D  & \text{  for each }  m_t \in Leaf(\bar \tree(\TA)) \label{eq:deadline} 
\end{align}
\end{subequations}
where $Leaf(\bar \tree(\TA))$ is the set of leaf nodes of $\bar \tree(\TA)$, $\phi^m$ is the guard of a transition leaving $m(l)$ under input $a^m$, $m'$ is an $a^m$ successor of $m$ (as in Defn.~\ref{defn:expTreeSoln}-2-c), i.e., $(m(l), a^m, \lambda, \phi^m, m') \in \Delta$, $g(c)$ is as defined in~\eqref{eq:LP}, and 
\begin{align}\label{eq:transformTree}
M^\tree(x, m') &= M(x, \pi^\TA_{r \to m'(l)}, length(\pi^\TA_{r \to m'(l)})).
\end{align}
As in~\eqref{eq:LP} and~\eqref{eq:transform}, each clock $x$ is mapped to sum of the delay values since its last reset based on the path from the initial location to the position of the constraint via $M^\tree(x, m')$~\eqref{eq:transformTree}. With a slight abuse of notation, $length(\pi^\TA_{r \to m(l)})$ is used to denote the index of the last transition along the path $\pi^\TA_{r \to m'}$ (automaton path obtained from the tree path from root $r$ to $m'$). Furthermore, the indices in~\eqref{eq:transform} are considered as relative indices in $\pi^\TA_{r \to m'}$ and assumed to map to $\{d_m \mid m \in Int(\bar \tree(\TA))\}$ in order not to complicate the notation. Note that $\phi^m$ is uniquely defined by Assumption~\ref{assm:samephi}. Essentially, the tree represents several paths. The delay variables are associated with the tree nodes and they are shared among the paths. If this MILP is feasible, then each of these paths is realizable via the corresponding delay sequence.  On the other hand, if the MILP~\eqref{eq:LP} defined for a path is not feasible, then the tree MILP~\eqref{eq:LPTree} can not be feasible. This property is exploited in Sec.~\ref{sec:mainold}. Finally, even if the MILPs~\eqref{eq:LP} defined for the tree paths are all feasible, the tree MILP might not be feasible.
Next, we define a control strategy $\mathcal{C}(\cdot)$ from a feasible solution of this MILP, and prove that $\mathcal{C}(\cdot)$ and $\gamma$ obtained from MILP~\eqref{eq:LPTree} solves Prob.~\ref{prob:main}.

\begin{proposition}\label{prop:strategy} Let $\TA=(L, l_0, \Sigma, C, \Delta)$ be a parametric timed automaton with parameter set $P$, $\{I_p\}_{p \in P}$ be the set of parameter ranges, and $\bar \tree(\TA)$ be a proper sub-tree of $\TA$ with respect to specification $(L_T, L_A, D)$. Let MILP~\eqref{eq:LPTree} be feasible for $\bar \tree(\TA)$, and $d^\star_m$ for each $m \in \bar Int(\tree(\TA))$, $\gamma^\star_p$ for each $p \in P$ be a solution, and let control strategy $\mathcal{C}$ w.r.t. $\bar \tree(\TA)$ and $d^\star_m \in \bar \tree(\TA)$ be defined as: 
\begin{align}\label{eq:strategy}
& \mathcal{C}(\pi = l_0, e_1, \ldots e_n, l_n) = \\
& \quad \begin{cases} 
(d^\star_m, a^m)     & \text{ if } \exists m \in  Int( \bar \tree(\TA) \ s.t. \pi^\TA_{r \to m} = \pi \nonumber \\
(\bot, \infty) & \text{ otherwise }
\end{cases}
\end{align}
Then, each $\rho \in \RUNS{\TA_\mathcal{C}(\gamma^\star)}$ satisfies $(L_T, L_A, D)$.
\end{proposition}

For the given automaton path $\pi$, the control strategy generates the delay and control action pair $(d^\star_m, a^m)$ associated with the last node $m$ of the corresponding tree path $\pi_{r \to m}$ ($\pi^\TA_{r \to m} = \pi$). Note that the strategy $\mathcal{C}(\cdot)$~\eqref{eq:strategy} is defined until the target set is reached due to the particular reachability specification. As this proposition highlights, a proper sub-tree of the exploration tree characterize a family of solutions by assigning an input to finite paths identified in the tree. Then, the solution of the MILP defines a strategy (as in~\eqref{eq:strategy}) by simultaneously finding parameter valuations for $\TA$ and delay values. 

\begin{example}~\label{ex:treeMILP}
The exploration tree $\tree(\TA)$ of the TA given in Fig.~\ref{fig:example} is finite and it has two proper sub-trees $\bar \tree_1(\TA)$ and $\bar \tree_2(\TA)$ such that $\bar \tree_1(\TA, r, a) \neq \emptyset$ (assigns $a$ to $l_0$) and $\bar \tree_2(\TA, r, b) \neq \emptyset$ (assigns $b$ to $l_0$). Note that no proper sub-tree assigns input $d$ to a node $m$ with $m(l) = l_a$ since $Post(l_a, d) \cap L_A \neq \emptyset$. The MILP constructed for $\bar \tree_2(\TA)$ is infeasible. In particular, $\bar \tree_2(\TA)$ includes $\pi_2$ from Ex.~\ref{ex:pathMILP} and the MILP~\eqref{eq:LP} defined for $\pi_2$ is infeasible, which is sufficient for infeasibility of the tree MILP.  On the other hand, the MILP~\eqref{eq:LPTree} defined for $\bar \tree_1(\TA)$ is feasible ($\gamma_{p_1} = 3, \gamma_{p_2} = 3$). $\bar \tree_1(\TA)$ includes 4 paths that end in $\{l_t\}$: $\pi_3: l_0, l_a', l_a, l_c', l_c, l_t$, $\pi_4: l_0, l_a', l_a, l_c, l_t$, $\pi_5: l_0, l_a, l_c', l_c, l_t$, $\pi_6: l_0, l_a, l_c, l_t$. The resulting strategy as defined in Prop.~\ref{prop:strategy} is (edges are omitted from the paths in $C(\cdot)$):
\begin{align*}
& C(l_0) = (0, a) , C(l_0, l_a') =   (3, a),  C(l_0, l_a) =   (3, c),    \\ 
& C(l_0, l_a', l_a) =  (3, c),  C(l_0, l_a, l_c') =   (3, c),   C(l_0, l_a, l_c) =   (9, c), \\
& C(l_0, l_a', l_a, l_c') =  (3, c),   C(l_0, l_a', l_a, l_c) = (9,c), \\  
& C(l_0, l_a, l_c', l_c) =   (6, c), C(l_0, l_a', l_a, l_c', l_c) =  (6, t),
\end{align*}
\end{example}

\begin{remark}\label{remark:samephi} For a TA violating  Assumption~\ref{assm:samephi}, a strategy can be computed by considering all guards associated with the location and control input in~\eqref{eq:guardtree}. In particular, consider location $l_s$ and input $a$  such that $(l_s, a, \lambda, \phi, l_t),  (l_s, a, \lambda', \phi', l_t')  \in \Delta$, with  $\phi \neq \phi'$.  Adding a constraint as in~\eqref{eq:guardtree} to the MILP for each inequality from $\phi \wedge \phi'$ guarantees that each symbolic path encoded in the tree (Defn.~\ref{defn:expTreeSoln}-c)) will be realizable when the MILP is feasible. 
\end{remark}

\section{Synthesis Algorithms}\label{sec:mainold}

In this section, we present an iterative method to construct the exploration tree as in Defn.~\ref{defn:expTree}, and a control strategy via a proper sub-tree (Defn.~\ref{defn:expTreeSoln}) as shown in Prop.~\ref{prop:strategy}. The method is summarized in Alg.~\ref{algo:soln}. 
The algorithm starts with the initialization of the root node (line 1) and expands the tree recursively by analyzing the input symbols and the corresponding transitions in a depth-first manner (described in Alg.~\ref{algo:tree}). For each considered input symbol, the feasibility of the corresponding automaton path is checked via MILP~\eqref{eq:LP} (line~\ref{line:isfeasible} of Alg.~\ref{algo:tree}). Thus, the exploration only continues through promising directions.  
Once the exploration tree construction terminates,  MILP~\eqref{eq:LPTree} is solved for each proper subtree until a feasible solution is found (lines~\ref{line:loop_ps_start}-\ref{line:loop_ps_end} of Alg.~\ref{algo:soln}).

  \algnewcommand{\IIf}[1]{\State\algorithmicif\ #1\ \algorithmicthen}
  \algnewcommand{\IIfe}[2]{\State\algorithmicif\ #1\ \algorithmicthen#2\ \algorithmicelse}
  \algdef{S}[FOR]{ForEach}[1]{\algorithmicforeach\ #1\ \algorithmicdo}

\begin{algorithm}[h!]
\caption{Synthesis($\TA$, $\mathcal{P}$, $(L_T, L_A, D)$ )} \label{algo:soln} 
\begin{algorithmic}[1] 
\Require  A PTA $\TA= (L, l_0, \Sigma, C, \Delta )$ with parameter set $P$, $\mathcal{P} = \{I_p \mid p \in P\}$ parameter intervals for each $p \in P$, specification $(L_T, L_A, D)$.
\Ensure Control strategy $\mathcal{C}$ and parameter valuation $\gamma$ such that each run from $\RUNS{\mathcal{C}(\mathcal{A}(\gamma))}$ satisfies $(L_T, L_A, D)$.
\State Initialize root $r$ of $\tree(\TA)$ with $r(l) = l_0$.~\label{line:root} %$Ch \leftarrow \emptyset, ps \leftarrow 0, \Sigma^{ps} \leftarrow \emptyset)$~\label{line:root}. 
\State $ps = $ ForwardAnalysis($r$, $r$, $\TA$, $P$, $\mathcal{P}$, $(L_T, L_A, D)$)\label{line:forward}
\ForEach{$i \in \{1, \ldots, ps\}$ }\Comment{Enumerate each posible solution.}\label{line:loop_ps_start}
\State $\bar \tree_i(\TA)= GetSolutionTree(i)$ \label{line:getSoln}
\State $\mathbf{d}, \gamma , \textit{soln} = Synthesis(\bar \tree_i(\TA))$\label{line:synthesis}
\IIf{$\textit{soln}$} {\Return $\mathcal{C}, \gamma = Controller(\bar \tree_i(\TA), \mathbf{d}, \gamma)$}\label{line:mapping}
\EndFor\label{line:forloopend}\label{line:loop_ps_end}
\State \Return No Solution \label{line:no_solution}
\end{algorithmic}
\end{algorithm}

\newcommand{\veryshortarrow}[1][3pt]{\tiny{\leftarrow}}
   
\begin{algorithm}[h!]
\caption{ForwardAnalysis($r$, $m$, $\TA$, $\mathcal{P}$, $(L_T, L_A, D)$)} \label{algo:tree} 
\begin{algorithmic}[1] % enter the algorithmic environment
    \Require $r$ is the root node, $m$ is a node reachable from $r$, $\TA$, $P$, $\mathcal{P}$, and $(L_T, L_A, D)$  are as in Alg.~\ref{algo:soln}. 
    \Ensure Construct tree, and return the number of possible proper trees that include $m$. 
    \IIf{$m(l) \in L_T$}{ \Return 1} \label{line:target}
    \IIf{$m(l) \in L_A$}{ \Return 0} \label{line:avoid}
    \State $ps^m = 0$ \Comment{The number of candidate solutions for $m$.}
    \ForEach{$a \in \Sigma(m(l))$} \Comment{For each admissible action.}\label{line:eachinput}
     	\If{$IsFeasible(root-m-a)$}~\label{line:isfeasible}
     		\State $ps = 1$ 
		\ForEach{$l' \in  Post(m(l),a)$} \label{line:postl1}
    	 		\State Create $m'$ with $m'(l) = l'$
     			\State Set  $\tree(\TA, m, a)  = \tree(\TA, m, a) \cup \{m'\}$ 
     			\State $ps = ps \times ${ForwardAnalysis$(root,m', \TA, \mathcal{P}, S)$}\label{line:recursive}
     		\EndFor\label{line:postl2}
     		\If{$ps == 0$} \Comment{No soln. from input $a$}~\label{line:infeasible_child}
     			\State Delete $\tree(\TA, m, a)$ \Comment{Remove each sub-tree.}
		\Else
			\State  $ps^m = ps^m + ps$  
			    		 \EndIf 
	\EndIf
     \EndFor\label{line:actionloop}
     \State \Return $ps^m$
\end{algorithmic}
\end{algorithm}

The forward analysis method (Alg.~\ref{algo:tree}) takes an exploration tree node $m$ as input, constructs the sub-tree rooted at $m$ recursively, and returns the number of different sub-trees that can be part of a proper sub-tree (a candidate solution Defn.~\ref{defn:expTreeSoln}) through $m$. It can be regarded as the number of different candidate solutions that contain $m$. 
Reaching a location from the target set (line~\ref{line:target}) or from the avoid set (line~\ref{line:avoid}) terminates the recursive construction. Otherwise, each admissible input is considered for the node (line~\ref{line:eachinput}). First, the feasibility of the timed automaton path induced by the exploration tree path from root to $m$ and input $a$ (line~\ref{line:isfeasible}) is checked via MILP~\eqref{eq:LP} from Prop.~\ref{prop:milp} (e.g. considering a location $l' \in Post(m(l),a)$ as the final location of the path). If this MILP is not feasible, i.e., the path is not realizable by any parameter valuation, the corresponding sub-trees of the exploration tree  ($m' \in \tree(\TA, m, a)$) are not constructed. On the other hand, if it is feasible, the exploration continues for each $l' \in Post(m(l), a)$ recursively (lines~\ref{line:postl1}-\ref{line:postl2}). 

The number of candidate solutions (proper sub-trees) associated with node $m$ and input $a$, denoted by $ps$, is the product of the number of solutions associated with the $a$-successors of $m$, i.e. $ ps = \Pi_{m' \in \tree(\TA, m, a)} ps^{m'}$. Note that each combination of these alternative choices can yield a different proper sub-tree of $\tree(\TA)$. Furthermore, if $ps^{m'}$ is $0$ for a node $m' \in  \tree(\TA, m, a)$, then the specification is not satisfiable through $m'$. As $\TA$ can reach $m'(l)$ non-deterministically when $a$ is applied at $m$, $ps$ is also set to $0$, and  each sub-tree associated with $m' \in \tree(\TA, m, a)$ is removed (line~\ref{line:infeasible_child}). Otherwise, the number of possible solutions through $m$ is incremented by $ps$ reflecting the sub-trees assigning $a$ to $m$.

A sub-tree constructed by Alg.~\ref{algo:tree} (extracted in line~\ref{line:getSoln} of Alg.~\ref{algo:soln}) satisfies conditions of Defn.~\ref{defn:expTreeSoln}. The first condition (1) follows from the initialization in line 1 of Alg.~\ref{algo:soln}. The condition that a node of the sub-tree belongs to the exploration tree (e.g. cond. (2)) trivially holds since nodes are added via $Post(m(l), a)$ relation (line~\ref{line:postl1}). The first base condition (line~\ref{line:target}) ensures that a child node is not constructed for a node $m$ when $m(l) \in L_T$ (2-b). The second base condition (line~\ref{line:avoid}) ensures that $m(l) \not \in L_A$ for any $m \in \bar \tree_{i}(\TA)$ since nodes with $0$ number of possible solutions are removed (see line~\ref{line:recursive} and~\ref{line:infeasible_child}) (2-a).  The connectivity (2-d) and the control assignment (2-c) conditions are satisfied by the enumeration performed with respect to the number of possible proper sub-trees ($ps$).

Note that since $\TA$ is non-deterministic, the feasibility analysis performed for paths (line~\ref{line:isfeasible} of Alg.~\ref{algo:tree}) is not sufficient to generate a control strategy. However, as the specification requires each run to satisfy the property, it is sufficient to prune violating runs. In particular, the feasibility of MILP from~\eqref{eq:LP} is a necessary condition for the feasibility of the MILP~\eqref{eq:LPTree} of the proper sub-trees that contain the path.
 Alg.~\ref{algo:tree} returns the number $ps$ of the proper subtrees of $\tree(\TA)$ that pass the path based feasibility check. In Alg.~\ref{algo:soln}, each proper sub-tree $\bar \tree_i(\TA)$ is extracted (line ~\ref{line:getSoln}),  MILP~\eqref{eq:LPTree} for the tree $\bar \tree_i(\TA)$ is solved (line~\ref{line:synthesis}), and if this MILP is feasible, a control strategy $\mathcal{C}(\cdot)$ as in~\eqref{eq:strategy} w.r.t. the MILP solution is returned.  By Prop.~\ref{prop:strategy}, we conclude that a strategy generated by Alg.~\ref{algo:soln} solves Prop.~\ref{prob:main}. 

Alg.~\ref{algo:soln} exhaustively searches all possible strategies via Alg.~\ref{algo:tree}. 
 Thus, if Alg.~\ref{algo:soln} reaches line~\ref{line:no_solution}, then a solution to Prop.~\ref{prob:main} does not exists. Consequently, when the algorithm terminates, either a strategy and parameter valuation pair solving Prop.~\ref{prob:main} is generated or a solution does not exist. 
 A final possibility is that the algorithm might not terminate. In particular, if $\TA$ has a loop, then $\tree(\TA)$ is infinite, and in this case Alg.~\ref{algo:tree} might fail to terminate. Next, we state an assumption that avoids zeno behavior by guaranteeing that the time progresses at each cycle ($l_i = l_j$ on a path): 
\begin{assumption}~\label{asm:zeno} For a TA $\TA = (L, l_0, \Sigma, C, \Delta )$, if an infinite run $\pi: l_0, e_1, l_1, e_2, \ldots$ is realizable by a delay sequence $d_0, d_1, \ldots,$ then for a positive constant $\epsilon$: \[ \text{if } l_i = l_j, j > i \text{ then } d_i + d_{i+1} \ldots d_{j-1} > \epsilon \]
\end{assumption}
Finally, we can guarantee that Alg.~\ref{algo:soln} finds a solution when one exists if timed automata $\TA$ satisfies Assumption~\ref{asm:zeno}. By the well-known pigeon hole principle, a path of length $| L | \cdot k$ includes a location at least $k$ times. By Assumption~\ref{asm:zeno}, if such a path is realizable, then the total duration of the corresponding delay variables are lower bounded by $k \cdot \epsilon$. Thus, the length of a path induced by the exploration tree path is upper bounded by $\frac{D}{\epsilon}$, as otherwise the resulting MILP~\eqref{eq:LP} is infeasible due to the time bound $D$. Consequently, if Assumption~\ref{asm:zeno} holds, the depth of the tree generated by Alg.~\ref{algo:tree} is bounded and the synthesis algorithm always terminates.

\begin{example} We run Alg.~\ref{algo:soln} on the TA $\TA$ introduced in Ex.~\ref{ex:running}.  As shown in Ex.~\ref{ex:treeMILP},  path $\pi_2$ is infeasible. Thus, $ps$ is set to $0$ for root $r$ and input $b$.  
In addition, $ps=0$ is assigned to trees with $\tree(\TA, m, d)  \neq \emptyset$ in line~\ref{line:avoid}.  As MILPs~\eqref{eq:LP} defined for paths $\pi_3,\pi_4,\pi_5$ and $\pi_6$ are feasible, $ps=1$ in Alg.~\ref{algo:tree} (line~\ref{line:forward}). As illustrated in Ex.~\ref{ex:treeMILP}, the corresponding MILP is feasible and results in a control strategy solving Prob.~\ref{prob:main}.
\end{example}

\section{Conclusion}
In this paper, we studied the controller and parameter synthesis problem for a PTA under unavoidability specifications with a deadline. We presented the candidate solutions symbolically with sub-trees of the exploration tree, and developed an algorithm to generate such trees. The algorithm is based on depth-first analysis and it uses an iterative feasibility check to terminate the exploration along infeasible directions. Finally, we presented an MILP based method to generate a feedback control strategy and a parameter valuation pair from a sub-tree such that the resulting TA satisfies the given specification.

\bibliography{ebru_ecc} 
\bibliographystyle{ieeetr}

\end{document}